\documentclass[10pt,onecolumn,english]{book}

\usepackage{geometry}
\usepackage{fancyhdr}
\usepackage{subfigure}
\usepackage{graphicx}
\usepackage{url}

\usepackage[english]{babel}
\usepackage{amsmath,amssymb,amscd,latexsym,dsfont}
\usepackage{cite}
\usepackage{textcomp}
\usepackage{psfrag}
\usepackage{rotating}
\usepackage{enumerate}
\usepackage{pdflscape}

\usepackage[final]{pdfpages}
\usepackage{xspace}
\usepackage{relsize}

\usepackage{url}
\usepackage{IEEEtrantools}

\usepackage{microtype}

\usepackage{todonotes}

\def\phdThesis{1}

\newcommand{\currentyear}{2023}
\newcommand{\authorname}{Mazen Mohamad}
\newcommand{\mytitle}{Understanding, Implementing, and Supporting Security Assurance Cases in Safety-Critical Domains}
\newcommand{\division}{Interaction Design and Software Engineering}
\newcommand{\techReportNumber}{XXXX}

\usepackage[T1]{fontenc}
\usepackage{graphicx}
\usepackage{todonotes}
\usepackage{afterpage}
\usepackage{multirow}
\usepackage{booktabs}
\usepackage{graphicx,subfigure}

\usepackage{url}
\usepackage{pgfplots}
\usepackage{float}
\usepackage{longtable}
\usepackage{multicol}
\usepackage[T1]{fontenc}
\usepackage{graphicx}
\usepackage{afterpage}
\usepackage{multirow}
\usepackage{booktabs}
\usepackage{comment}
\usepackage{url}
\usepackage{pgfplots}
\pgfplotsset{compat=1.17}
\usepackage{float}
\usepackage{longtable}
\usepackage{multicol}
\usepackage{tabularx}
\usepackage{siunitx}
\usepackage[linewidth=1pt]{mdframed}
\newcolumntype{s}{>{\hsize=.4\hsize}X}
\newcolumntype{P}[1]{>{\raggedright\arraybackslash}p{#1}}
\usepackage{listings}
\usepackage{array,multirow,graphicx}
\usepackage{caption}
\usepackage[numbers]{natbib}
\newrobustcmd*{\citefirstlastauthor}{\AtNextCite{\DeclareNameAlias{labelname}{given-family}}\citeauthor}
\usepackage{colortbl}
\usepackage{adjustbox}
\usepackage{xr}

\usepackage{todonotes}
\usepackage[utf8]{inputenc}
\usepackage[T1]{fontenc}
\usepackage{microtype}
\usepackage{csquotes}
\usepackage{booktabs}
\usepackage{enumitem}
\usepackage{tikz} 
\usepackage{pgfplots}
\usepackage{tablefootnote}
\usepackage{threeparttable}
\pgfplotsset{compat=1.14}
\usepackage{xspace}
\def\signed#1{{\leavevmode\unskip\nobreak\hfil\penalty50\hskip2em
		\hbox{}\nobreak\hfil\raise-3pt\hbox{#1}
		\parfillskip=0pt \finalhyphendemerits=0 \endgraf}}

\newcommand{\inlinequote}[1]{``\emph{#1}''}

\usetikzlibrary{shapes,arrows,positioning}

\tikzset{decision/.style={diamond, draw, fill=yellow!20, 
    text width=6em, text badly centered, node distance=3cm, inner sep=0pt},
block/.style={rectangle, draw, fill=gray!20, 
    text width=12em, text centered, rounded corners, minimum height=4em},
small_block/.style={circle, draw, fill=white!20, 
    text width=0em, text centered, rounded corners, minimum height=0em},    
line/.style={draw, -latex'},
cloud/.style={draw, ellipse,fill=red!20, node distance=3cm,
    minimum height=2em},}
\usepackage{tikz-dependency}

\lstdefinestyle{mystyle}{
    keepspaces=true,                 
    numbers=left,                    
    numbersep=5pt,                  
    showspaces=false,                
    showstringspaces=false,
    showtabs=false
}
\usepackage{longtable}
\usepackage{lscape}
\newcommand{\phdISBNNumber}{978-91-8069-329-5}
\newcommand{\phdSeriesNumber}{xxxx}
\newenvironment{widequotation}{\list{}{\listparindent 1.5em \itemindent\listparindent
		\rightmargin 0pt \parsep 0pt plus 1pt}\item\relax}
{\endlist}
\usepackage{xspace}
\def\signed#1{{\leavevmode\unskip\nobreak\hfil\penalty50\hskip2em
		\hbox{}\nobreak\hfil\raise-3pt\hbox{#1}
		\parfillskip=0pt \finalhyphendemerits=0 \endgraf}}
\newsavebox\mybox

\begin{document}
\frontmatter

\ifx\phdThesis\undefined
\newcommand{\degreetitle}{Doctor of Philosophy}
\newcommand{\reportno}{TODO}
\newcommand{\reportNoText}{Technical Report No \techReportNumber \\
 ISSN 1652-876X\\ }
\else
\newcommand{\degreetitle}{Doctor of Philosophy}
\newcommand{\reportNoText}{ISBN \phdISBNNumber\\
Doktorsavhandlingar vid Chalmers tekniska h\"{o}gskola, Ny serie nr \phdSeriesNumber .\\
ISSN 0346-718X\\ 
\vspace{1cm}

\noindent
Technical Report No \techReportNumber \\
}
\fi


\thispagestyle{empty} 
\begin{center}
  \textsc{Thesis for The Degree of \degreetitle}\\
\end{center}

\vspace{6cm}
\begin{center} \Large \mytitle
\end{center}

\vspace{1cm}
\begin{center}
\textsc{\authorname} \\
\end{center}

\vspace{2cm}
\begin{figure}[h]
  \begin{center}
     \includegraphics[width=30mm]{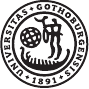}
  \end{center}
\end{figure}
\vspace{2cm}

\begin{center}
Division of \division \\
Department of Computer Science \& Engineering\\
Chalmers University of Technology and University of Gothenburg\\
Gothenburg, Sweden, \currentyear \\
\end{center}

\newpage
\thispagestyle{plain}

\vspace{2cm} \noindent \textbf
\mytitle\\

\noindent
\textsc{\authorname}\\

\vfill

\noindent
Copyright \copyright \currentyear \space \authorname \\
except where otherwise stated. \\
All rights reserved.\\ 

\noindent
ISBN: 978-91-8069-329-5 (PRINT)\\
ISBN: 978-91-8069-330-1 (PDF)\\

\noindent 
Department of Computer Science \& Engineering\\
Division of \division \\
Chalmers University of Technology and University of Gothenburg\\
Gothenburg, Sweden\\

\noindent
Cover:\\
The cover picture illustrates a summary of the main contributions of this thesis work.\\
\vskip 0.4cm
\noindent This thesis has been prepared using \LaTeX.

\noindent
Printed by Chalmers Reproservice,\\
Gothenburg, Sweden \currentyear.


\newpage
\begin{table}
  \begin{flushright}
	\large{\emph{``The only truly secure system is one that is powered off, cast in a block of concrete, and sealed in a lead-lined room with armed guards.'' \\ - Gene Spafford}}  
\end{flushright}
\end{table}

\newpage
\thispagestyle{empty}

\cleardoublepage \addcontentsline{toc}{chapter}{Abstract}
\thispagestyle{empty}
\section*{Abstract}
  The increasing demand for connectivity in safety-critical domains has made security assurance a crucial consideration.  In safety-critical industry, software, and connectivity have become integral to meeting market expectations. Regulatory bodies now require security assurance cases (SAC) to verify compliance, as demonstrated in ISO/SAE-21434 for automotive. However, existing approaches for creating SACs do not adequately address industry-specific constraints and requirements.  
 In this thesis, we present CASCADE, an approach for creating SACs that aligns with ISO/SAE-21434 and integrates quality assurance measures. CASCADE is developed based on insights from industry needs and a systematic literature review. We explore various factors driving SAC adoption, both internal and external to companies in safety-critical domains, and identify gaps in the existing literature. 
 Our approach addresses these gaps and focuses on asset-driven methodology and quality assurance. We provide an illustrative example and evaluate CASCADE's suitability and scalability in an automotive OEM. We evaluate the generalizability of CASCADE in the medical domain, highlighting its benefits and necessary adaptations.
 Furthermore, we support the creation and management of SACs by developing a machine-learning model to classify security-related requirements and investigating the management of security evidence. We identify deficiencies in evidence management practices and propose potential areas for automation. Finally, our work contributes to the advancement of security assurance practices and provides practical support for practitioners in creating and managing SACs.

\vspace{5 mm} \noindent{\textbf{Keywords:}}  

\vspace{3 mm} \noindent{Security, Assurance case, Safety-critical, Automotive systems, Arguments, \\Evidence, Security claims}

\newpage
\thispagestyle{empty}

\cleardoublepage \addcontentsline{toc}{chapter}{Acknowledgement}
\chapter*{Acknowledgment}
\vspace{5 mm}

First of all, I would like to express my sincere gratitude to my supervisors Riccardo Scandariato, Jan-Philipp Steghöfer, and Eric Knauss for all the support, advice, patience, collaboration, and trust they provided me during this journey.  I also thank my examiner Andrei Sabelfeld for trusting me and giving me the freedom to conduct my research. To my former examiner, Ivica Crnkovic, even though you are no longer with us, your memory and impact on my life will always be cherished. May your soul rest in peace.\\

Thank you to all my colleagues and friends at the Interaction Design and Software Engineering division, Amna, Babu, Bea, Cristy, Habib, Hamdi, Hazem, Joel, Krishna, Linda, Malsha, Rebekka, Ricardo, Rodi, Sushant, Teodor, and Wardah, and all faculty members in the interaction design and software engineering division for creating a very nice working environment and for all the great social and sports activities. \\

I would like to extend a special thanks to my dear friends Ranim, Razan, and Khaled for all the support and encouragement. I cannot express enough gratitude for your willingness to lend your time and effort to support me at all times. \\

I would also like to thank my industrial partners at Volvo and Volvo Cars for all their support.\\ 

Finally, I wish to express my deepest gratitude to my parents, friends, and my wife Rim, who did not spare a chance to motivate me and provide me with love and encouragement. I also wish to thank my children Bassam for making me smile every day, and Sol for shining my life. \\

This work is partially supported by the CASUS research project funded by VINNOVA, a Swedish funding agency. 

\newpage
\thispagestyle{empty}


\cleardoublepage \addcontentsline{toc}{chapter}{List of Publications}
\chapter*{List of Publications}
\label{A:Papers}

\section*{Appended publications}
This thesis is based on the following publications:
\renewcommand{\labelenumi}{[\Alph{enumi}]}
\begin{enumerate}

\item M. Mohamad, A. Åström, Ö. Askerdal, J. Borg, R. Scandariato
\newblock ``Security Assurance Cases for Road Vehicles: an Industry Perspective''\\
\newblock {\em Proceedings of the 15th International Conference on Availability, Reliability and Security, 2020} \cite{includedA-1}.\\
\newblock {\em Extended version is available on https://arxiv.org/abs/2003.14106} \cite{includedA-2}.

\item M. Mohamad, J.P. Steghöfer, R. Scandariato
\newblock ``Security Assurance Cases – State of the Art of an Emerging Approach''\\
\newblock {\em Empirical Software Engineering Journal 26 (4), 70, 2021} \cite{includedB}.

\item M. Mohamad, R. Jolak, Ö. Askerdal, J.P. Steghöfer, R. Scandariato
\newblock ``CASCADE: An Asset-driven Approach to Build Security Assurance Cases for Automotive Systems''\\
\newblock {\em ACM Transactions on Cyber-Physical Systems 7 (1), 1-26, 2023} \cite{includedC}.

\item M. Fransson, A. Andersson, M. Mohamad, J.P. Steghöfer
\newblock ``Security Assurance Cases in the Medical Domain: A Case Study''\\
\newblock {\em Proceedings of the 50th Euromicro Conference on Software Engineering and Advanced Applications (SEAA), 2024 } \cite{includedD}.\\

\item  M. Mohamad, JP. Steghöfer, A. Åström, R. Scandariato
\newblock ``Identifying security-related requirements in regulatory documents based on cross-project classification''\\
\newblock {\em Proceedings of the 18th International Conference on Predictive Models and Data Analytics in Software Engineering, 2022} \cite{includedE}.\\

\item  M. Mohamad, JP. Steghöfer, E. Knauss, R. Scandariato
\newblock ``Managing Security Evidence in Safety-Critical Organizations''\\
\newblock {\em Journal of Systems and Software 214, 112082, 2024} \cite{includedF}.

\end{enumerate}

\newpage
\section*{Other publications}
The following publications were published before or during my PhD studies, or are currently in submission/under revision.
However, they are not appended to this thesis, due to contents overlapping that of appended publications or contents not related to the thesis.

\renewcommand{\labelenumi}{[\alph{enumi}]}
\begin{enumerate}

  \item R. Jolak, T. Rosenstatter, M. Mohamad, K. Strandberg, B. Sangchoolie, N. Nowdehi, R. Scandariato
 \newblock ``CONSERVE: A Framework for the Selection of Techniques for Monitoring Containers Security''\\
 \newblock {\em Journal of Systems and Software 186, 111158, 2022} \cite{othera}.

\item H.P. Samoaa, A. Longa, M. Mohamad, M.H. Chehreghani, P. Leitner
 \newblock ``TEP-GNN: Accurate Execution Time Prediction of Functional Tests Using Graph Neural Networks''\\
 \newblock {\em Product-Focused Software Process Improvement: 23rd International Conference, PROFES 2022, Jyväskylä, Finland, November 21–23, 2022, Proceedings, 2022} \cite{otherb}.

   \item J.P. Steghöfer, B. Koopmann, J.S. Becker, M. Törnlund, Y. Ibrahim, M. Mohamad
 \newblock ``Design Decisions in the Construction of Traceability Information Models for Safe Automotive Systems''\\
 \newblock {\em 2021 IEEE 29th International Requirements Engineering Conference (RE), 2021} \cite{otherc}.

\item  M. Mohamad
 \newblock ``Towards Understanding and Applying Security Assurance Cases for Automotive Systems''\\
 \newblock {\em Licentiate thesis - Chalmers Library, 2021} \cite{otherd}.

\item M. Mohamad, Ö. Askerdal, R. Jolak, J.P. Steghöfer, R. Scandariato
\newblock ``Asset-driven Security Assurance Cases with Built-in Quality Assurance''\\
\newblock {\em 2021 IEEE/ACM 2nd International Workshop on Engineering and Cybersecurity of Critical Systems (EnCyCriS), 2021} \cite{othere}.

 \item M. Mohamad, G. Liebel, E. Knauss
 \newblock ``LoCo CoCo: Automatically constructing coordination and communication networks from model-based systems engineering data''\\
 \newblock {\em Information and Software Technology Journal 92, 179-193, 2017} \cite{otherf}.

  \end{enumerate}

\cleardoublepage \addcontentsline{toc}{chapter}{Personal Contribution}
\thispagestyle{empty}
\section*{Research Contribution}

To define my contribution to the appended papers in this thesis, I use the CRediT (Contribution Roles Taxonomy) model, defined by Brand et al. [1].

\begin{table*}[hb!]
\caption{Contributions of Mazen Mohamad to the appended papers of this thesis}
\label{tbl:contributions}
\begin{tabular}{lP{0.75cm}P{0.75cm}P{0.75cm}P{0.75cm}P{0.75cm}P{0.75cm}}
\toprule
             
        \textbf{Role / Paper}  &  \textbf{A} & \textbf{B}       & \textbf{C}  &   \textbf{D}  &   \textbf{E}  & \textbf{F}  \\
        \midrule
Conceptualization & \checkmark & \checkmark& \checkmark& \checkmark& \checkmark & \checkmark  \\  

Methodology & \checkmark & \checkmark& \checkmark& \checkmark& \checkmark& \checkmark   \\  

Software & & & & & \checkmark &   \\  
Validation &\checkmark & \checkmark&\checkmark & & \checkmark& \checkmark  \\  
Formal analysis & \checkmark & \checkmark &  \checkmark&  &  \checkmark&  \checkmark \\  
Investigation & \checkmark& \checkmark& \checkmark& & \checkmark & \checkmark \\

Resources &  & \checkmark &  & \checkmark & \checkmark & \checkmark  \\
Data Curation & \checkmark & \checkmark & \checkmark &\checkmark  & \checkmark & \checkmark  \\
Writing - Original Draft & \checkmark& \checkmark& \checkmark& & \checkmark &\checkmark  \\
Writing - Review \& Editing &\checkmark & \checkmark& \checkmark& \checkmark& \checkmark & \checkmark \\
Visualization &\checkmark& \checkmark& \checkmark& & \checkmark&  \checkmark \\
Supervision & & & &\checkmark &  &  \\
Project administration &\checkmark & \checkmark&\checkmark &\checkmark & \checkmark & \checkmark \\
Funding acquisition & & & & &  &  \\
 
\bottomrule
\end{tabular}
\end{table*}

\tableofcontents 

 
\mainmatter

\chapter{Introduction}
\label{chap_Introduction}
%
\label{kappa:thesisAtGlance}
With the growing demand for connectivity in the products and services provided by companies in safety-critical domains, security has become an increasingly important consideration. In the automotive industry, for example, software now plays a major role in vehicles, and connectivity is vital to satisfy market expectations for features and services such as mobile phone integration and navigation systems.

This brings the matter of security assurance to the forefront, prompting the question of how to ensure and demonstrate the security of a product. This concern is particularly significant in complex systems, comprising multiple sub-systems and dependencies among them and involving numerous stakeholders, such as various suppliers for different components.

Regulatory and standardization bodies started demanding a systematic approach to ensure the security of products and processes in safety-critical domains. Therefore, Security Assurance Cases (SAC) have been mandated in various documents as a means of verifying security compliance, such as  ISO/SAE-21434 \cite{iso21434} for the automotive domain. Consequently, companies in safety-critical domains have begun exploring methods for creating and sustaining these cases, as well as integrating them into their existing operational practices.

SACs are designed to argue and prove that a certain artifact, e.g., product, user functionality, or component, is acceptably secure. The term \emph{acceptably secure} can be interpreted in different ways. Hence, a SAC needs to be complemented with contextual information for defining what level of security shall be achieved in order to determine if the artifact in question is acceptably secure. In safety-critical domains, the market demand for the level of security is usually driven by regulatory requirements and best practice reports. However, different companies might have their own internal definition of acceptably secure based on their security policies and would then incorporate that in their SACs.

Assurance cases have been applied for many years in safety-critical domains, such as avionics and automotive. Many companies have dealt with cases designed for safety, known as safety cases, which are mandatory for ensuring functional safety in road vehicles according to ISO-26262 \cite{iso26262}. This presents an opportunity for knowledge transfer from the safety domain to the security domain. Nevertheless, this transfer of knowledge must be approached with care, taking into account the distinctions between the two domains.

In this work, we provide support for practitioners to create SAC. We provide CASCADE, an approach for creating SAC which takes into consideration the gaps between the state-of-the-art and the industrial needs and is compliant with the requirements of ISO/SAE-21434. We also provide support for the creation and management of the main components of a SAC, which are the arguments and evidence.

CASCADE is based on insights from two studies: one done in collaboration with two large Original Equipment Manufacturers (OEM) about the needs and drivers of work in SAC in the automotive industry and one systematic literature review in which we identified gaps between the industrial needs and the state of the art.

As an initial step, we conducted an exploration of various factors that drive the adoption of security cases in the automotive domain. We classify the drivers into two main categories. \emph{Internal drivers} that focus on understanding the requirements and needs originating from automotive companies. Thirteen different usage scenarios were identified where SAC can be applied. These scenarios spanned the entire lifecycle of automotive products and involved diverse roles within automotive companies. Furthermore, these scenarios imposed specific requirements for SAC, e.g., quality assurance, in order to make them useful for industrial applications. \emph{External drivers}, which focus on the requirements from external entities, e.g., regulators and standardization bodies. These play a crucial role in identifying the constraints of how SACs should look like. These drivers were identified by analyzing how SACs were referenced in various documents, including regulations, standards, and best practices, within the major automotive markets such as the EU, US, and China. We identified thirteen documents that explicitly or implicitly required SAC or suggested their usage to fulfill specific document requirements.
Moreover, we conducted a systematic literature review to understand if these needs are covered in the literature and identify potential gaps. We systematically examined different characteristics such as usage scenarios, approaches for creating SAC, and available tool support.
Upon analyzing our findings, several noteworthy observations emerged. Firstly, we identified a wide range of approaches for creating SAC in the literature, but none of them adequately addressed the specific constraints and requirements of the industry. Additionally, there was a notable absence of quality assurance measures for SAC in the reviewed literature. Although the literature provides multiple potential usages of SACs, we observed that they do not consider all the internal needs and suggested applications identified within automotive companies.

Considering these factors, we developed our own approach to creating SACs, drawing insights from both the industry needs and the existing literature. The approach focused on two key aspects:

\begin{itemize}
    \item Alignment with ISO/SAE-21434: We ensured that our approach aligned with the requirements and work products outlined in the ISO/SAE-21434 standard. This alignment aimed to establish a coherent relationship between the standard and the resulting SAC.
    \item Integration of Quality Assurance: An important aspect of our approach was the integration of quality assurance within the SAC themselves. We recognized the significance of ensuring the quality and reliability of the cases as an intrinsic part of the creation process.
\end{itemize}

Our developed approach, called CASCADE, follows an asset-driven methodology for creating security assurance cases,  while also incorporating built-in quality assurance measures. To demonstrate the application of CASCADE we provided an illustrative example use case derived from ISO/SAE-21434. Furthermore, we conducted an evaluation of CASCADE in collaboration with security experts from a prominent automotive OEM. The evaluation results demonstrated the suitability of CASCADE for integration into industrial product development processes. The elements and principles of CASCADE aligned effectively with the working methodologies and practices within the company. Moreover, CASCADE exhibited the potential to scale and cater to the diverse requirements and needs of the organization, given its large-scale operations and the complexity of its products.

To assess the generalizability of CASCADE beyond the original domain in which it was created, i.e., automotive, we conducted an evaluation of CASCADE in the context of a large medical device manufacturer with an established agile development workflow. We investigated the regulatory context as well as the adaptations needed in the development
process. Through this evaluation, we identified areas where adopting CASCADE could bring benefits to the medical domain and also areas that may require modifications for effective implementation.
Throughout our investigation, we identified a total of 17 use cases in which a SAC serves to address both internal and external needs. 
We established a connection to safety assurance by incorporating relevant information from the risk assessment matrix into the SAC. To facilitate the integration into the development process, we proposed the introduction of a new role and guidelines for design reviews and production releases. Additionally, we recommended the inclusion of supplementary criteria for the definition of done.

Moreover. we investigated ways to support the creation and management of SACs. We supported the creation of the argumentation part of the SAC (in which claims about the security of the system are made) by creating a machine-learning model to classify security-related requirements. This would enable practitioners to quickly identify requirements in deployed systems that would be the base of the argumentation. For new projects, the model assists in identifying those requirements that need to be considered for a SAC. Additionally, it helps the practitioners to identify security-related requirements in regulatory documents for compliance reasons. 
The evaluation of the model indicates the feasibility of identifying security requirements when trained on heterogeneous data sets including specifications from multiple
domains and in different styles. It also shows the ability of such a classifier to identify security requirements in real-life regulations.

To address the management of the evidence, which is the other main part of a SAC, we conducted an investigation into how safety-critical organizations handle security evidence. Our study examined the current practices of security evidence management, the integration of security evidence into an organization's development process, existing procedures for managing security evidence, challenges in this context, and the potential application of automation to support practitioners in evidence management.

Based on our findings, it is evident that there is a deficiency in the maturity level of managing the increasing demands associated with security evidence management. Our research revealed that companies primarily tackle the development of security evidence at the team level, lacking an overarching organizational framework. Interestingly, we discovered that the challenges related to security evidence management are primarily organizational in nature rather than technical. Additionally, we identified specific areas where the implementation of automation could enhance the efficiency of evidence management processes.


\section{Research Focus}
\label{rq:ResGoal}
This research is motivated by the growing recognition of the importance of SAC in various safety-critical domains, e.g., automotive.
The primary goal of this research is ``\emph{to support practitioners in safety-critical domains to prove that their systems and products are acceptably secure with the help of security assurance cases}''.  


To achieve this overall goal, we addressed the following goals in this licentiate thesis:
\begin{itemize}
  \item \textbf{Goal 1:} Identify the gaps between the current state of the art of SAC as presented in the literature and the specific needs of safety-critical domains.
    \item \textbf{Goal 2:} Develop and assess an approach for creating SAC that considers the identified gaps.
    \item \textbf{Goal 3:} Assist practitioners in implementing SACs by supporting the generation and management of its main components.
\end{itemize}

By addressing the first goal, we gain an understanding of the market demand of the safety-critical sector for security assurance and whether the literature offers sufficient solutions to meet these demands. The second goal helps address the potential gaps between the literature and the industrial needs and provides an approach for practitioners to reason about and increase confidence in the security of their systems. Hence, it would help them prove that their systems and products are secure enough to meet the market demand. The third goal addresses supporting the practitioners in creating these cases and also managing the different artifacts that emerge from the assurance process. This helps maintain the assurance cases and sustain the security confidence they provide.



To reach the goals of this thesis, we formulate the following research questions:

\begin{LaTeXdescription}
 \item [RQ1:] What are the drivers for working with security assurance cases in safety-critical domains?
 
 This question addresses the emergence of several standards and regulations that are forcing industries to develop a methodology for SAC in order to stay compliant and avoid litigation risks. We call these the \emph{external drivers} that will impose constraints on what SAC should look like. 
  The need to develop a strategy for SAC is also perceived by the automotive companies as an opportunity to improve their cybersecurity development process. As such, the question also takes up the \emph{internal drivers} related to this aspect.
  
 \item[RQ2:] What are the gaps in the state of the art when it comes to the industrial applicability of SAC? 
 
 This question aims at identifying gaps in the state of the art with respect to the needs of companies in the automotive domain from two perspectives:
 \begin{itemize}
     \item Approaches for the creation of SAC
     \item Support to assist the practitioners in creating SAC
 \end{itemize}
  
 \item [RQ3:] How can an approach for the construction of security assurance cases fulfill the needs of the automotive domain?
 
 The purpose of this question is to investigate how an approach for SAC creation can be built in order to fulfill both the external and internal needs of automotive companies, as well as close the gaps between research and the industrial needs for SAC adoption.
\item [RQ4:] To what extent can an approach for creating SAC in automotive be applied in other safety-critical domains?

This question aims at studying the generalizability of a SAC creation approach built for a specific safety-critical domain into other similar domains. In particular, this studies the potential of using SAC in the medical domain and we study how CASCADE, which is an approach for creating SAC built in the automotive domain, can be applied in the medical domain.

\item [RQ5:] To what extent can the creation of security arguments be supported by a machine learning model for predicting security requirements?

In this question, the possibility of using a machine learning classifier that classifies whether a certain requirement is security related or not to support the creation of security arguments is studied. We also study the possibility of using the classifier for real-life regulatory documents.

\item [RQ6:] How can security evidence be managed to support the creation of SACs?

As security evidence is a crucial part of an assurance case, this question aims at studying how security artifacts should be managed in safety-critical organizations in order for them to be used as security evidence in SAC.
\end{LaTeXdescription}

By answering RQ1 and RQ2 we address Goal 1, while answering RQ3, RQ4 addresses Goal 2, and Goal 3 is addressed by answering RQ5 and RQ6.
\section{Background}

In this section, we provide background about security assurance cases and approaches to create them. We also discuss some related work.
\subsection{Security Assurance Cases}
\begin{figure}[htb]
        \centering
        \includegraphics[width=\textwidth]{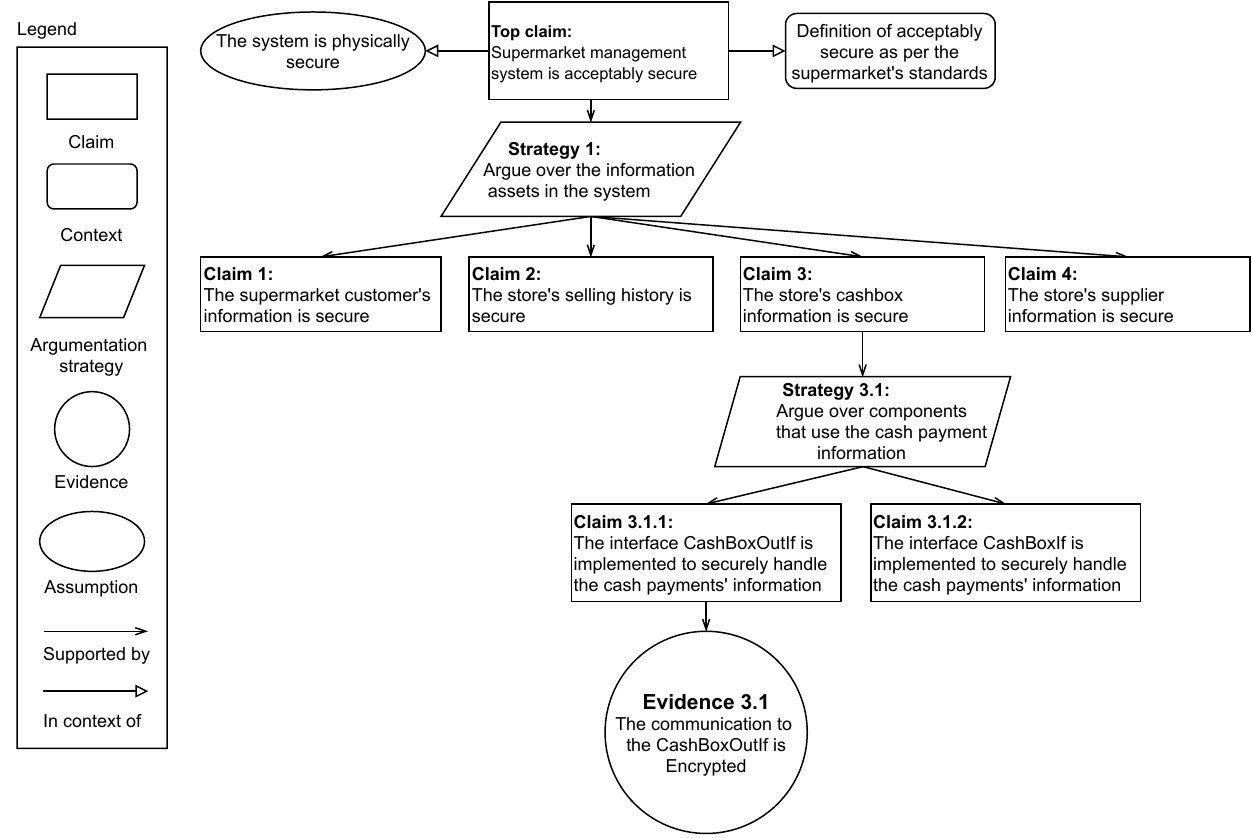}
        \caption{An example of a security assurance case}
        \label{fig:kappasacExample}
    \end{figure}
    


According to the GSN standard \cite{GSN_standard}, an assurance case is: \inlinequote{A reasoned and compelling argument, supported by a body of evidence, that a system, service or organization will operate as intended for a defined application in a defined environment.}

 Assurance cases can be represented visually or textually. 
 Security assurance cases consist of two main components. These are the argumentation and the evidence. 
 
 Argumentation is a process used to reason about things in a systematic way to support a certain case.  There are different types and models of arguments, e.g., the Toulmin theory \cite{arguments1}, which has a structure that is similar to SACs but has limitations when it comes to complex cases \cite{arguments2}.
 Each claim in an assurance case needs to be verified and supported by evidence, which distinguishes it from the argumentation types proposed in the literature \cite{arguments3}.

 The evidence part of an assurance case includes artifacts that support the claims made in the argumentation.
 
 Figure \ref{fig:kappasacExample} displays an illustration of an assurance case documented with the GSN notation. This particular example is a section of a broader case developed for a supermarket management system.
The SAC in Figure \ref{fig:kappasacExample} consists of several nodes: claim (also called goal), context, strategy, assumption (also called justification), and evidence (also called solution). The high-level claim is usually located at the top of the case and is called the top claim. It is broken down into sub-claims based on specific strategies. The claims describe the objectives that need to be ensured in the case, such as the preservation of a particular property.
 An example of a strategy is breaking a claim down based on the information assets of the system in question as shown in \emph{Strategy 1} in Figure \ref{fig:kappasacExample}. 
 Claims are broken down iteratively until they reach a point where evidence can be assigned to justify them. Examples of evidence include test results, monitoring reports, and code review reports. The assumptions made while applying the strategies are explicitly stated using the assumption nodes. For instance, this may involve assuming that the system in question is not physically accessible to threat agents. Finally, the scope of a claim is set using the context nodes. An example of a context is the definition of an acceptably secure system.

 Assurance cases have a well-established usage in various domains for safety-critical systems \cite{bloomfield2010}. Examples of these domains include the automotive industry, where safety cases have been utilized to demonstrate compliance with the functional safety standard ISO 26262~\cite{palin2011,birch2013,iso26262}, and the medical field, where safety cases have been employed to ensure the safety of medical devices \cite{medicalSafetyExample}. 
 However, there is a growing interest in utilizing these cases for security purposes as well. The automotive industry, for instance, has a new requirement in the  standard for road vehicles ISO 21434~\cite{iso21434}, which necessitates the establishment of cyber-security arguments. 

\subsection{Approaches for SAC creation}
 This section provides an overview of various approaches for creating SAC that are relevant to this thesis work. In particular, we review these approaches in relevance to our approach CASCADE which is the main contribution of this thesis. CASCADE is an asset-driven approach that facilitates the creation of SAC while incorporating quality assurance measures. The approach is influenced by the ISO/SAE-21434 cybersecurity standard for the automotive industry. Consequently, we review papers that employ asset-based argumentation strategies, leverage security standards, or are conducted in the automotive domain.

\subsubsection{Asset-based approaches}


 There has been interest in asset-based approaches to develop the argument part of SAC in research. Assets are defined as valuable artifacts for a specific organization, project, or system. Several studies have explored the use of asset decomposition as a strategy to break down claims in SAC.

 For instance, Biao et al.~\cite{20_xu2017} propose dividing the argument into multiple layers, each with a different pattern. Assets are treated as one of these layers, and the pattern used to create it includes claims that the assets are "under protection" and strategies to break down critical assets. However, Biao et al.'s approach only focuses on creating arguments and does not address the evidence part or the quality of the cases, unlike our work with CASCADE.

 Luburic et al.~\cite{luburic2018} have proposed an asset-based approach for security assurance that utilizes three types of information: \emph{(i)} asset inventories; \emph{(ii)} Data Flow Diagrams (DFD) for assets and the components handling them; and \emph{(iii)} the security policy governing the components. 
 Their approach centers around assets, which are linked to security goals, and the argument focuses on protecting the assets throughout their life-cycle by securing the components that store, process, and transmit them. The provided SAC is of a high-level and features two strategies: "reasonable protection for all sensitive assets" and an argument for the data-flow of each relevant component. 
 The authors demonstrate their approach using an example conference management system, noting that their limitations include asset and data flow granularity. In contrast, our work also utilizes assets as a basis for the approach, but we extend the argument to concrete security requirements. We draw our strategies from an industrial standard and validate our approach in collaboration with an OEM. Additionally, we expand our approach to include quality aspects of the case.

\subsubsection{Standard-based approaches}


 Several studies have used standards as a basis for creating SAC arguments. However, no previous work has specifically addressed the upcoming ISO/SAE-21434 standard for cybersecurity in the automotive industry. For instance, Finnegan et al. \cite{8_finnegan2014,45_finnegan2014} proposed a security case framework for medical device security assurance that integrates various standards and best practices to develop a comprehensive security argument pattern. 

 Ankrum et al. \cite{9_ankum2005} investigated how requirements from safety-critical domain standards can be mapped to assurance cases using GSN and ASCAD notations, including the Common Criteria for Information Technology Security Evaluation, ISO/IEC 15408:1999 \cite{iso15408}. However, they faced challenges and drew lessons learned from the mapping process.

 In contrast, our work focuses on using the ISO/SAE-21434 standard to structure our approach for creating SAC, while also addressing the specific needs of the automotive industry. We aim to provide concrete security requirements and strategies based on the standard, and consider the quality of the cases as well.

\subsubsection{SAC in safety-critical industries}
T
 There have been few studies evaluating the use of SAC in  safety-critical domains.
 
 In automotive in a study by Cheah et al. \cite{39_cheah2018}, the authors discuss the challenges of introducing a security engineering process in the automotive domain and propose a classification approach to security test results using severity ratings. The study includes two case studies, one involving a Bluetooth connection to the infotainment system of a vehicle and the other involving an aftermarket diagnostics tool. 
 The results of both studies provide severity-rated evidence that could be used to prioritize countermeasure development and add evidence to security assurance cases. Although no security assurance case is created in this study, the authors suggest that the severity-rated evidence could be used as evidence in a security assurance case.

 In the medical domain Arnab Ray and Rance \cite{ray_security_2015} suggest using security assurance cases (SACs) to enhance the security and safety of medical devices. They propose integrating SACs into the design, implementation, verification, and documentation of medical devices to promote desirable security practices. This approach contrasts with the current trend of creating assurance cases after development to meet regulatory requirements. 


\section{Methodology}


This section summarizes the research methodologies applied to answer the research questions of this thesis. 

\begin{table*}
\caption{The research methodologies used to answer the research questions of this thesis}
\label{tbl:methodologies}
\begin{tabular}{P{7.6cm}P{2.2cm}P{1cm}}
\toprule
\textbf{Research question} & \textbf{Methodology} & \textbf{Paper}\\
\midrule
\textbf{RQ1:} What are the drivers for working with security assurance cases in safety-critical domains?& Case Study & A,D,F \\  \midrule

\textbf{RQ2:} What are the gaps in the state of the art when it comes to the industrial applicability of SAC?& SLR & A,B \\
 \midrule  
\textbf{RQ3:} How can an approach for the construction of security assurance cases fulfill the needs of the automotive domain?&  DSR & A,B,C\\
 \midrule
\textbf{RQ4:} To what extent can an approach for creating SAC in automotive be applied in other safety-critical domains?& Case Study & C,D\\
 \midrule
\textbf{RQ5:} To what extent can the creation of security arguments be supported by a machine learning model for predicting security requirements? &  Experiment &E\\
\midrule
 \textbf{RQ6:} How can security evidence be managed to support the creation of SACs? & Case Study & F\\
\bottomrule
\end{tabular}
\end{table*}

Table \ref{tbl:methodologies} shows the research methodologies used to answer each of the research questions along with the relevant appended papers. RQ 1, RQ 4, and RQ 6 were answered through case studies with various qualitative research methods. RQ 3 was addressed using the design science research method, RQ 2 through a systematic literature review, and RQ 5 through experimentation.

\subsection{Case studies with qualitative research methods}
\begin{figure}[htb]
        \centering
        \includegraphics[width=\textwidth,trim={0 5cm 0 2cm},clip]{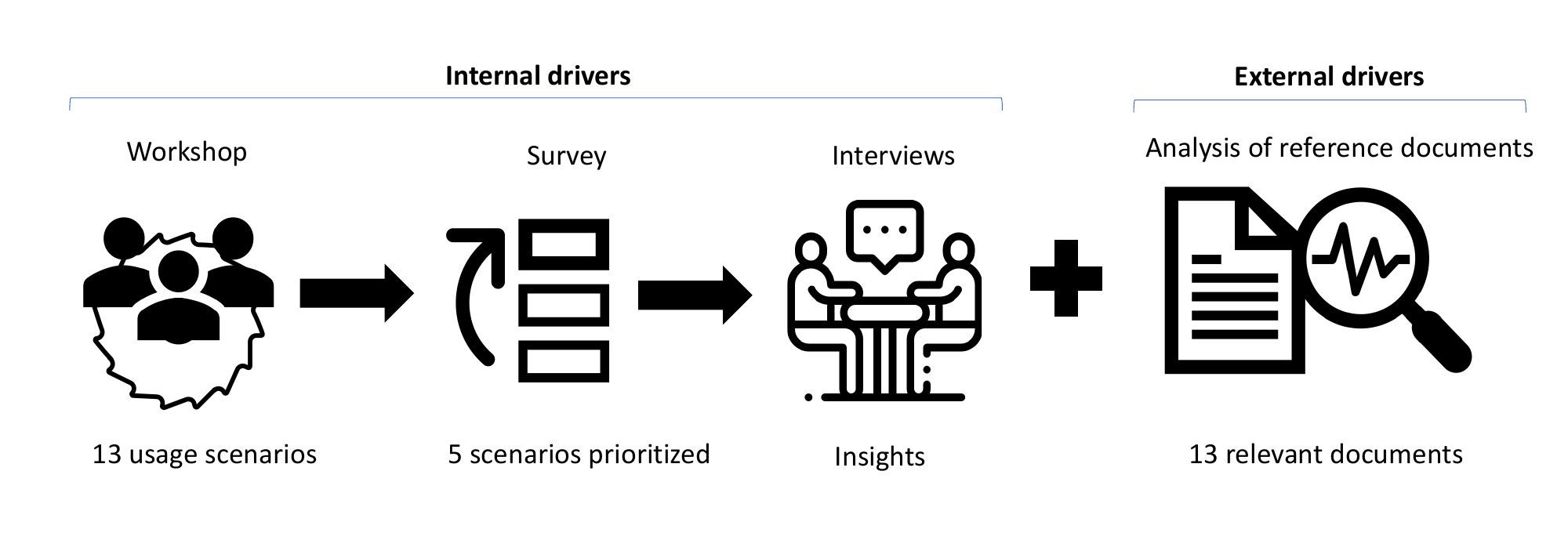}
        \caption{Qualitative research methods -- Paper A}
        \label{fig:ws}
    \end{figure}

\begin{figure}[tb]
\centering
\includegraphics[width=\textwidth]{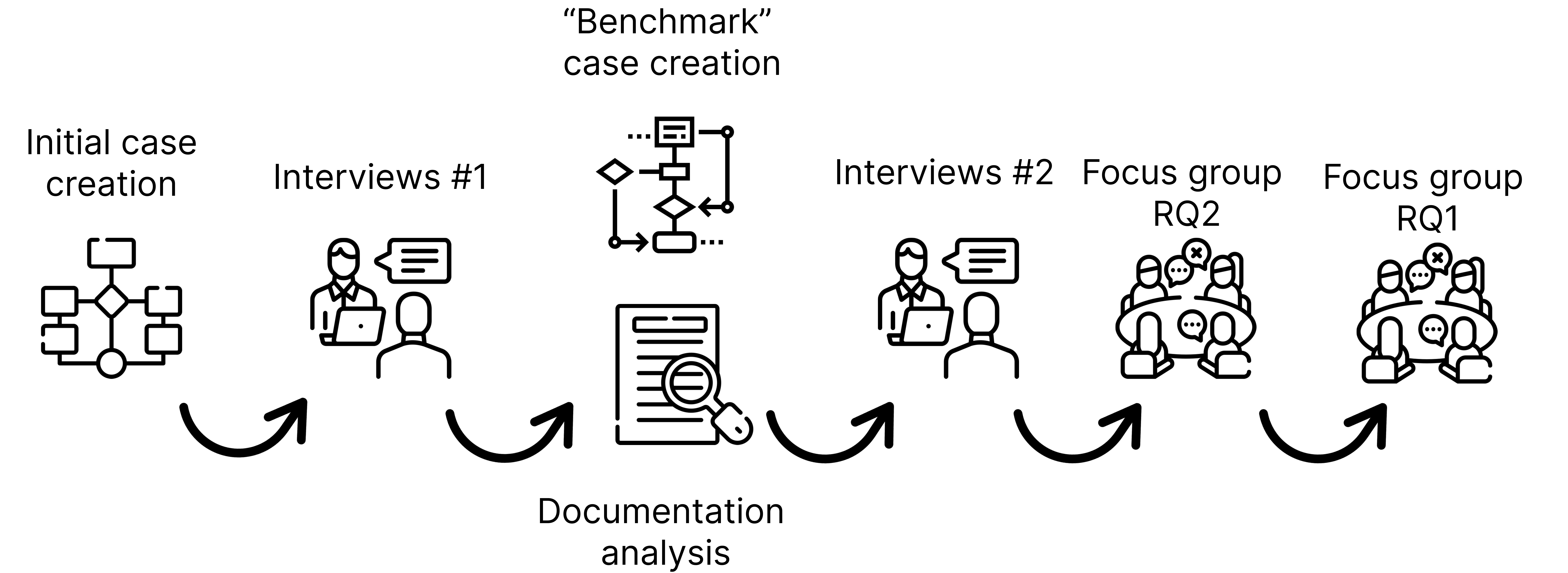}
\caption{The methods used for the case study in chronological order -- Paper D}
\label{fig:ws2}
\end{figure}

We used various qualitative research methods in Paper A and Paper D to answer RQ 1 and RQ 4, as shown in Figures \ref{fig:ws} and \ref{fig:ws2} respectively. These include a workshop, a survey, focus groups, and one-to-one interviews.
Additionally, Focus groups were used in Paper F to collect data from six different case companies. 
The reason we chose these methods is that we wanted to get an in-depth analysis of SAC usage and evidence management in a real-world context. Additionally, we wanted to make use of the flexibility that qualitative research methods provide in terms of exploration and data collection, as the topic we are studying is complex and relatively new to the targeted subjects. 

\subsubsection{Workshop}
In paper A, at a workshop with a large automotive OEM, we invited stakeholders from various backgrounds and divided them into three groups to brainstorm real-life usage scenarios for security assurance cases. Each group consisted of 4 participants with diverse roles and competencies, and we asked them to describe their ideas as user stories, such as ``As a <<role>> I would use security assurance cases for <<usage>>''~\cite{cohn2004}. From the participants' input, we compiled a set of distinct scenarios for the next step.
To analyze the results, we sent it out to 10 security experts from an automotive OEM, and asked them to select the top five scenarios by assigning a rank from 1 to 5 to them, where 5 is assigned to the most valuable scenario for the company.

\subsubsection{Interviews}
Interviews were used in Papers A to gain a better understanding of the most important scenarios and acquire diverse stakeholder perspectives, we prioritized and identified the top five scenarios from a security perspective. 
The prioritization was done by sending the scenarios collected from the workshop to 10 security experts from an automotive OEM and asking them to identify the top 5 scenarios with respect to the value they would provide for the company. 
We then selected key stakeholders for each scenario to interview in person based on the relevance of their roles to the actors of the user stories. For instance, the actor of one usage scenario is a \emph{legal risk owner}. Hence, the selected interviewee had the role \emph{senior legal counsel} in the company.
 Each interview was organized into four parts: 
 \begin{itemize}
     \item The \textbf{value} that SAC might bring to the stakeholder in terms of, e.g., efficiency, and quality management
     \item The interviewees' technical opinions on how the \textbf{Content and structure} of SAC should be, e.g., in terms of the level of detail and types of claims. 
     \item The \textbf{integration} of SACs with the current way of working, and whether it could fit in the current activities, or would require modifications to the process.
     \item The \textbf{challenges and opportunities} that the stakeholders foresee in implementing and using SACs in their contexts.
 \end{itemize}
 
We recorded and analyzed the interviews using deductive coding, and validated the results with the interviewees.

We also used interviews in Paper D, where we conducted two rounds of semi-structured interviews. In the first round, we interviewed key stakeholders to gather insights about regulations, requirements, and security demands in the context of the medical domain. Regulations from the medical domain require compliance with certain requirements for product marketing approval. We assessed if SAC created with CASCADE fulfills the requirements of regulation in the medical domain for product marketing approval. In our analysis, we investigated which parts of the documentation map to the building blocks of CASCADE and which requirements in the regulations they fulfill.
In the second round, we conducted interviews to collect qualitative information, assessing the compliance of CASCADE-created SACs with medical domain regulations.

\subsubsection{Focus groups}
Focus groups were used in Papers D and F.
In paper D, two focus groups were conducted. The first focused on discussing the maintainability process for CASCADE SACs in the medical domain, involving experts with different roles and responsibilities. The participants filled in a questionnaire, brainstormed workflows, and discussed their fits for key processes. The second focus group validated the results and discussed the applicability of CASCADE, with participants familiar with the system of interest. The researchers presented preliminary results, engaged in discussions, and created use cases of SAC at the case company.

In paper F, a total of seven focus group sessions were conducted, with varying numbers of practitioners involved. 
The focus group sessions aimed to understand the state of security-related work, security assurance practices, types and management of security evidence, logistics of evidence storage and access, usage of evidence, and automation of evidence management. 
Thematic coding was used to analyze and organize the data in Paper F. The process involved familiarizing with the data, creating a preliminary codebook, conducting coding based on pre-defined codes, refining the codebook, identifying emerging codes, and searching for patterns. Two workshops were held to analyze coded statements and cluster them into themes. A total of 56 clusters were identified, which were then used to draw insights and conclusions from the data.
\subsection{Systematic Literature Review (SLR)}
\begin{figure}[htb]
        \centering
        \includegraphics[width=\textwidth,trim={0 5cm 0 2cm},clip]{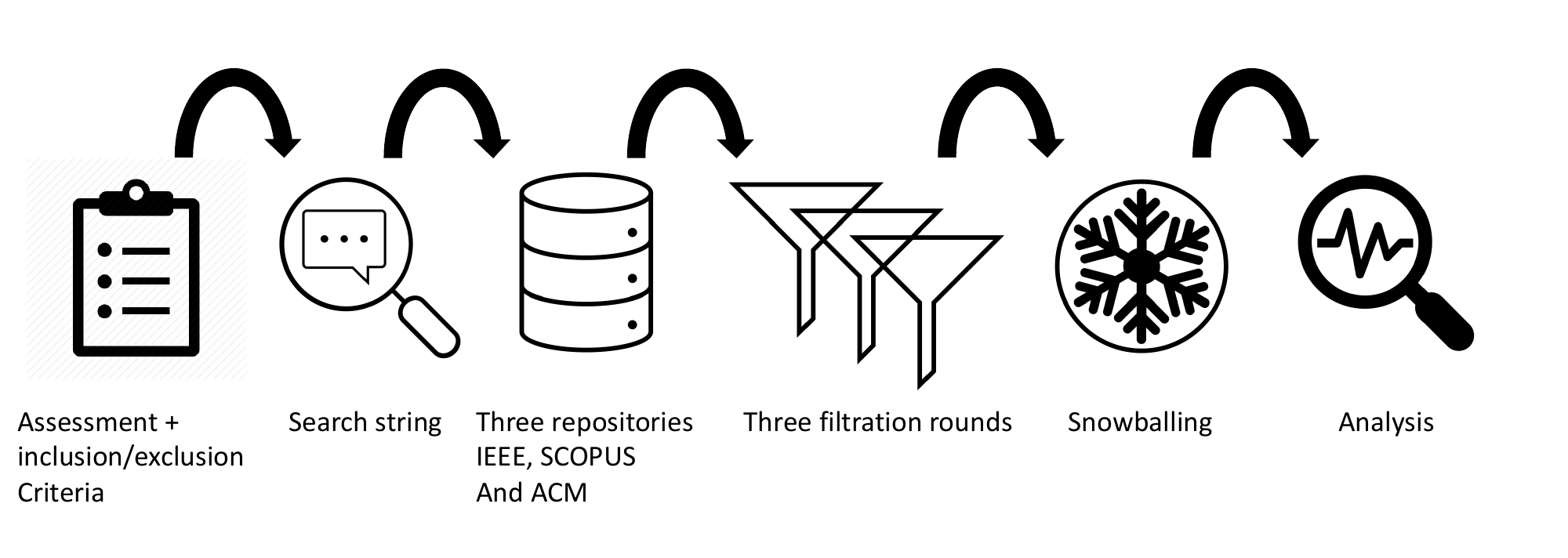}
        \caption{Systematic Literature Review steps -- Paper B}
        \label{fig:slr}
    \end{figure}

In Paper B, we conducted a Systematic Literature Review (SLR) to summarize and synthesize existing research on SACs. SLR was chosen as a research method because it helps to identify knowledge gaps, which is one of the main goals of this thesis work.
%
 We focused on approaches for creating security assurance cases, evidence of their validity, support for their adoption, and rationale for their adoption.
 
%
Following the guidelines by Kitchenham et al. \cite{kh2007guidelines}, we performed the SLR in six steps, as shown in Figure \ref{fig:slr}.


 We developed assessment and inclusion/exclusion criteria through collaborative brainstorming sessions involving the three authors.

To familiarize ourselves with the terminology used in security assurance, we did a manual search of relevant papers published in reputable venues within the last five years. Based on that, we created the search string which we executed on three libraries (IEEE Xplore, ACM Digital Library, and Scopus) and got a total of 8440 results.
%

 Next, we applied inclusion/exclusion criteria in three filtration rounds. The first round filtered based on title and keywords, resulting in 211 included studies. In the second round, we applied criteria to abstracts and conclusions, reducing the number to 49. Finally, we thoroughly reviewed these 49 papers, applying the criteria to the entire text, resulting in 44 included studies.

We conducted \emph{backward snowballing}~\cite{snowballing} by examining the references in the included papers and incorporating potential gray literature as well. This step added 7 papers (including 2 technical reports) to our review.

Finally, the included 51 included studies were analyzed based on the pre-defined assessment criteria to answer our research questions.


\subsection{Design Science Research (DSR)}

\begin{figure}[htb]
    \centering
    \includegraphics[width=\linewidth ,trim={1.3cm 2cm 2cm 1cm},clip]{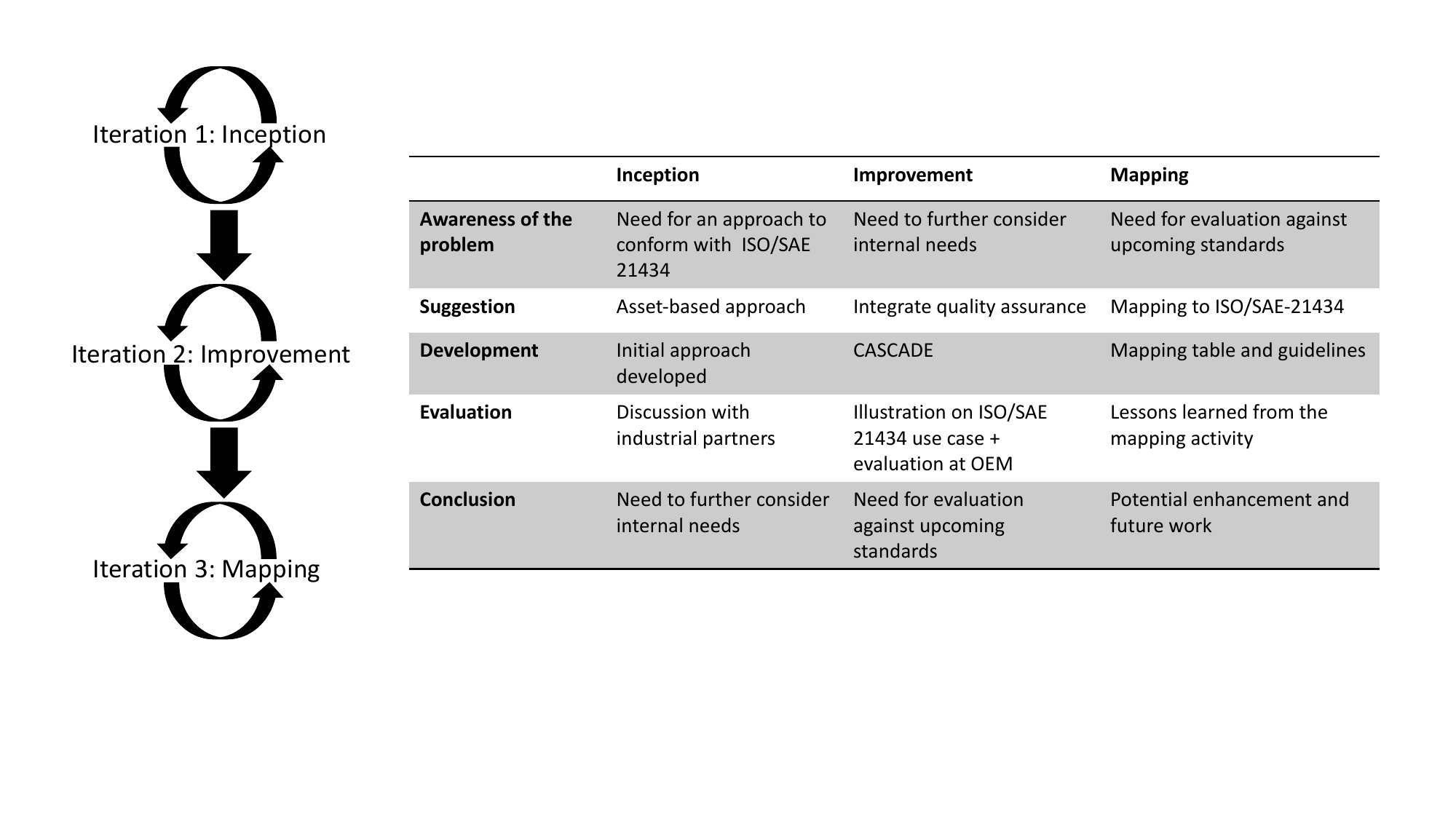}
    \caption{Three-iteration Design Science Research -- Paper C}
    \label{fig:dsr}
\end{figure}



In Paper C, we used Design science research, which is a problem-solving methodology, aiming at developing artifacts to extend existing boundaries in a given context \cite{Hevner2004}.

We conducted three research iterations following the design science guidelines proposed by Hevner et al.~\cite{Hevner2004} and the five-step process proposed by Vaishnavi and Kuechler~\cite{vaishnavi2004}, consisting of the \emph{awareness of the problem}, \emph{suggestion}, \emph{development}, \emph{evaluation} and \emph{conclusion} steps, as shown in Figure \ref{fig:dsr}.

The Inception iteration aimed to address the needs for security assurance cases previously identified in Paper A.
As a result of the iteration, we proposed an initial asset-based approach, and an online case for a supermarket system \cite{cocome} was used to illustrate it.
Feedback from security experts at automotive OEMs led to the identification of the need to align the approach with internal company processes and ensure quality assurance. 

 In the Improvement iteration, CASCADE, an asset-based approach for security assurance case creation with built-in quality assurance, was developed based on the initial approach and feedback received. CASCADE was evaluated using the example case of a headlamp item from ISO/SAE-21434 \cite{iso21434} and presenting the outcome to security experts at an OEM.
In conclusion, we identified areas for future enhancement of CASCADE to fulfill a wider range of the internal needs of the company.

 In the Mapping iteration, the requirements and work products of the ISO/SAE 21434 standard were mapped to CASCADE elements to evaluate its coverage and identify potential improvements. Two researchers independently conducted the mapping, reaching a 71\% agreement, and resolved disagreements through calibration exercises. 
 The mapping results and lessons learned were analyzed to propose potential enhancements for CASCADE. The study also provides a guideline for replicating the mapping activity and identifies areas for future work to improve CASCADE's applicability in the automotive domain.

\subsection{Experimentation}
An experimental setup was used in Paper E to create and test a classifier of security-related requirements. This methodology helps to test our hypothesis that machine learning can effectively classify these requirements in requirements specifications and real-life regulatory documents. 
 The study collected fifteen different requirement specifications from various sources, including commercial projects, student projects, domain-specific guidelines, industrial projects, and research projects, resulting in a total of 3,880 requirements. 
The included documents were selected based on accessibility and heterogeneity. 
The requirements were categorized into security and non-security classes based on predefined criteria. The labeling process involved manual labeling, revision of pre-labeled data, and identification of security requirements in partially labeled documents. Inter-rater agreement was calculated to ensure labeling consistency. 
 For the analysis, a cross-project prediction experiment was conducted using a Random Forest classifier. Feature extraction was performed using the Bag of Words representation with TF-IDF weighting. To address the class imbalance, the Synthetic Minority Oversampling Technique (SMOTE) was applied. 
 Additionally, the classifier was tested on three regulatory documents from the automotive domain to evaluate its performance in predicting security requirements in regulations. The labeling of the regulatory documents was done by a different person to mitigate bias. The same pre-processed dataset and algorithm used in the previous experiment were employed for this analysis.

\section{Contributions}
In this section, we provide a summary of the main contributions of each paper toward answering our research questions.

\subsection{RQ1: What are the drivers for working with security assurance cases in safety-critical domains?}
To answer this research question we conducted the studies presented in Papers A, D, and F.
In Paper A, we focus on the automotive domain, while Paper D is conducted on the medical domain, and Paper F in multiple domains including automotive and medical.
We explored the external drivers of working with SAC by analyzing the different regulations and standards where SAC are either explicitly required or where applying them would be beneficial to cover the requirements of the regulations and standards. 

We also explored the internal drivers of security assurance case (SAC) adoption.
Our results clearly demonstrate the potential value of implementing SAC within safety-critical companies. They highlight the wide range of stakeholders, such as product owners and compliance team members, who can utilize SAC for various purposes, including quality assessment and communication with suppliers, throughout all stages of a  product's life cycle, such as design and development.

\subsubsection{Drivers of working with SAC}
The driving force behind security assurance case (SAC) implementation in a safety-critical company lies in the value it can provide to individuals in various roles within the organization. Consequently, we have identified and prioritized 13 different usage scenarios for SAC in the automotive context. By assessing their potential added value, we have determined the top 5 scenarios, which are detailed in Table \ref{tbl:rq1:usageScenarios}. Similar scenarios were identified in the medical domain, however, with different roles.

\begin{table*}
\caption{Top 5 usage scenarios identified at an automotive company}
\label{tbl:rq1:usageScenarios}
\begin{tabular}{lP{10cm}}
\toprule
\textbf{US 2} & As a member of the compliance team, I would use detailed SAC to prove to authorities that the company has complied to a certain standard, legislation, etc., and show them evidence of my claim of compliance.  \\  
\midrule
\textbf{US 6} & As a product owner, I would use SAC to make an assessment of the quality of my product from a security perspective and make a road-map for future security development. \\
\midrule    
\textbf{US 12} &  As a legal risk owner, I would use SAC in court if a legal case is raised against the company for security related issues. I would use the SAC to prove that sufficient preventive actions were taken.\\
\midrule 
\textbf{US 8} & As a member of the purchase team, I would include SAC as a part of the contracts made with suppliers, in order to have evidence of the fulfillment of security requirements at delivery time, and to track progress during development time.\\
\midrule 
\textbf{US 3} &  As a project manager, I would use SAC to make sure that a project is ready from a security point of view to be closed and shipped to production.\\
 
\bottomrule
\end{tabular}
\end{table*}

Through interviews conducted in multiple companies in safety-critical domains, we extracted a set of drivers for companies wanting to adopt SAC in their work, as shown in Table \ref{tbl:rq1:recom}.

\begin{table*}
\small
\caption{Drivers of SAC work in safety-critical companies}
\label{tbl:rq1:recom}
\begin{tabular}{P{4cm}P{7cm}}
\toprule
\textbf{Driver} & \textbf{Description}\\
\midrule
The importance to cover both product and process to comply with regulations and standards & 
 Security-related standards and regulations encompass both process requirements for secure product development and post-release security measures.
\\  
\midrule
The need for SAC on whole products over sub-projects & 
 In industries with complex product development, such as automotive, products are typically organized into multiple projects, including delta projects for making changes. In such cases, it is recommended to create SAC at the product level rather than the project level.
\\
\midrule
Customer demands and requirements & Failing to meet customers' security requirements and demands could result in a loss of customers and economic damage to companies\\ 
\midrule

Mitigating the threat of litigation & If a security flaw is found in the products provided by the companies, then security assurance and evidence in particular can be employed to show what has been done and mitigate the risk of litigation.\\
\midrule

Essential that SAC work follows the development process &  
 SAC can be constructed for existing products, but it is crucial to integrate SAC work into the organization's development process moving forward.
\\
\midrule 
The need to actively assess the quality of SAC & 
 SACs have various purposes and criticality levels within the organization. To ensure appropriate usage, it is important to establish clear quality levels for each SAC.
\\
\midrule 
A common language is key to smooth collaboration with suppliers &  
 When collaborating with suppliers, SACs should be constructed in a format that allows for interchangeability. This facilitates the integration of suppliers' SACs with the corresponding product's SAC.
\\
\midrule 
The importance to plan for shared ownership with suppliers & 
 When suppliers need to keep certain parts of the SAC confidential, mechanisms should be in place to maintain overall SAC quality. This can be achieved through the use of a black box with meta information. Furthermore, the ownership of the entire case must be addressed, as a single stakeholder may not possess the complete SAC.
\\
\bottomrule
\end{tabular}
\end{table*}


\subsection{RQ2: What are the gaps in the state of the art
when it comes to the industrial applicability of SAC?}
In order to address this question, we must consider the industrial needs, as discussed in Paper A. Additionally, we need to examine the existing literature to identify any gaps. Paper B contributes by conducting a Systematic Literature Review that investigates various research questions related to the applicability of SAC in industry. This includes studying the motivations, approaches, validations, and support reported in the literature for SAC creation.

\subsubsection{Wide variety of approaches, but lacking sufficient coverage of industrial needs}
 Existing literature offers a diverse range of studies that examine various approaches to SAC creation, with a particular emphasis on the argumentation aspect. However, these approaches do not take into account the industry-specific requirements of companies, such as those in the automotive sector.

 The range of approaches available allows organizations to select those that align with their work processes and security artifacts. For instance, an agile-oriented company may opt for a SAC approach suited for iterative development~\cite{18_othmane2016}. However, the decision should consider constraints, benefits, and challenges associated with its adoption, such as the impact on the work process. Unfortunately, these aspects are not addressed in the literature, leaving the burden on the adopter.

The literature also lacks systematic assessments of approaches' effectiveness in achieving conformance with specific standards. Similarly, there is a scarcity of studies comparing approaches in different contexts. Consequently, organizations face the challenge of exploratory selection, which consumes significant time and resources.

Future studies should address the granularity level achievable or necessary with different approaches for creating SAC. It is crucial to consider the potential use cases for SAC and determine the appropriate level of granularity based on those. For instance, for companies outsourcing development work, would SAC created through the security assurance-driven software development approach \cite{34_vivas2011} be applicable? If so, at what level should these cases be created, such as the feature level or the complete product level?

\subsubsection{Lack of quality assurance}
 The quality assurance aspect is limited in the literature reviewed in Paper B. Three main concerns arise. Firstly, there is a scarcity of industrial involvement, possibly due to a lack of interest or challenges in obtaining relevant and sensitive security-related data for validation. Secondly, the creation and validation of SAC presented in the literature is predominantly conducted by the authors of the studies, which hinders the exploration of challenges and drawbacks in real-world applications. Addressing these limitations is essential to enhance the practical applicability of SAC.

 The generalizability of approaches regarding argumentation strategies is another concern. The reviewed approaches employ diverse argumentation strategies, such as threat analysis, requirements, or risk analysis. However, they lack validation and critical examination of whether these approaches are limited to the employed strategies or if they can accommodate other strategies. Future research should validate these approaches using various types of strategies.

 The absence of quality assurance mechanisms in SAC is another issue. It is crucial for the argumentation in SAC to be comprehensive and reliable for their usefulness. However, the literature lacks substantial coverage of this aspect, except for a few studies that have partially addressed it (e.g., \cite{1_cyra2007,11_chindamaikul2014,25_rodes2014}).
 The assessment of the evidence in SAC is equally important. This can be done by introducing metrics to evaluate the degree to which evidence supports a specific claim. The interrelationship between claims and evidence should be examined to determine if a claim is adequately supported by the assigned evidence.
 

\subsubsection{Imbalance in coverage}
\label{disc:coverage}
 The literature lacks coverage of various matters related to SAC. For instance, needs and drivers such as supplier management, SAC quality assurance, and organizational aspects are not sufficiently covered. This highlights a weakness in the proposed approaches, as SAC elements cannot be evaluated in isolation. For instance, when creating security arguments, it becomes challenging to determine the appropriate evidence to associate with them. Similarly, assessing the adequacy of claim justification based on evidence granularity is difficult. The same applies to the evidence aspect, as identifying the claims that the suggested evidence can support becomes problematic without considering the overall context of the SAC. The incompleteness of the approaches is evident as there are no links between studies focusing on different elements of SAC.

 Insufficient attention is given to the assessment and quality assurance of SAC in other areas, as discussed previously. Moreover, there is a lack of studies addressing the post-creation phase of SAC. To remain useful, SAC must be regularly updated and maintained throughout the life cycles of the targeted products and systems. This was emphasized in Paper A. Establishing traceability links between the created SAC and the associated artifacts is crucial. Although many SAC approaches employ Goal Structuring Notation (GSN), which allows referencing external artifacts through context and assumption nodes, these nodes are seldom utilized in the reviewed studies' examples.

%
 Additionally, there is a scarcity of studies focusing on the organizational aspects of SAC implementation, such as the ownership of SAC and the management of sub-cases when collaborating with suppliers.

\subsection{RQ3: How can an approach for the construction of security assurance cases
fulfill the needs of the automotive domain?}

 Papers A,B, and C contribute to addressing RQ3. Building on the insights from RQ1 and RQ2, we developed the CASCADE approach for SAC creation, which is the primary contribution of Paper C. CASCADE is an asset-driven approach with integrated quality assurance.

\subsubsection{Design goals of CASCADE}
 CASCADE draws inspiration from the automotive cybersecurity standard SAE/ISO-21434 \cite{iso21434}. This standard's conformance is recognized as a key driver for SAC work in Paper A, but none of the papers included in the SLR of Paper B addressed it.
CASCADE was designed to achieve the following goals:
\begin{itemize}
    \item Make assets the driving force of the SAC to allow creating security assurance based on what is valuable in the system. 
    \item Embed quality assurance in the approach to make sure the outcome satisfies the desired quality by the adopting entity.
    \item Divide the approach into different layers and blocks, so that different people can work on them in different development phases.
    \item Enable re-usability and scalability to prevent overhead and work repetition while creating SAC on lower-level items.
\end{itemize}
\subsubsection{Structure of the approach}
\begin{figure}[hbt!]
\begin{center}
  \includegraphics[width=0.9\linewidth]{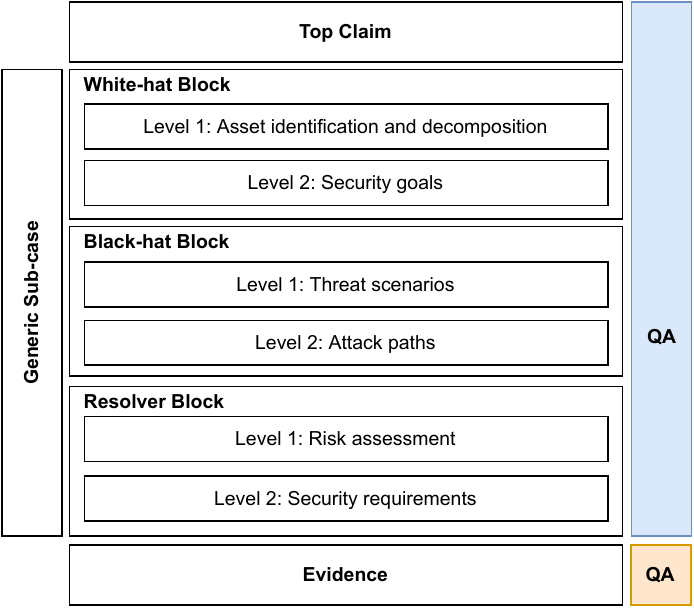}
  \caption{The CASCADE approach for creating security assurance cases} 
  \label{fig:kappa:assetApproach}
\end{center}
\end{figure}

CASCADE consists of blocks as shown in Figure \ref{fig:kappa:assetApproach}. These blocks correspond to the requirements and work products of SAE/ISO-21434. 

 The \textbf{Top Claim} within CASCADE represents the primary security claim associated with the specific artifact. It encompasses the claim itself, along with its contextual information and any assumptions made to define the claim's scope.
 The \textbf{Generic sub-case} in CASCADE facilitates reusability and scalability by incorporating a sub-case that extends beyond the specific artifact being addressed. This sub-case applies to a broader context, allowing for the utilization of common elements across multiple artifacts. For instance, if a company establishes a cybersecurity policy enforced by rules and processes, this policy can be employed in security claims for all its products.
 The \textbf{White-hat} block initiates with the identification of assets, aligning with our approach's design goals and priorities.
 Asset identification involves analyzing the system to identify potential targets of attacks. We establish connections between assets and the main claim by identifying the existing assets and the components that utilize or have access to them. Decomposing assets involves examining their types to determine if they impact a specific component or a larger span of the artifact in question.
 We also analyze the relationships between assets, such as the dependencies among them. To connect assets with the lower level of the approach, which is the security goals, we identify the corresponding security properties for each asset. We focus on the Confidentiality, Integrity, and Availability (CIA) triad. Once we have determined the relevant security properties for each asset, we formulate claims that represent the security goals which is defined as  preserving a security concern (CIA) for an asset \cite{secGoals}.

 In the \textbf{Black-hat} block, our focus is on identifying scenarios that could compromise the fulfillment of the security goals and pose a risk to our assets. Once we have formulated claims related to achieving the security goals, we proceed by identifying threat scenarios and creating claims to counteract these scenarios. We establish connections between these claims and the corresponding claims for achieving the security goals. Additionally, we identify potential attack paths that could lead to the realization of a threat scenario. For each threat scenario, there may be multiple associated attack paths. We then formulate claims that negate the possibility of these attack paths being realized.

 In the \textbf{Resolver block} of the CASCADE approach, we establish the connection between the claims derived from the attack paths and the corresponding evidence. At this stage, we assess the risk associated with the identified attack paths. Depending on the level of risk, the creators of the SAC formulate claims to address the risk by accepting, mitigating, or transferring it.
 Then the requirements of risk treatments identified in the previous level are expressed as claims. The level of detail in this block can vary based on the desired usage of the SAC. For example, if the SAC is intended for internal development team assessment, a fine-grained requirement decomposition may be needed, potentially down to the code level. On the other hand, if the SAC is meant for communication with external parties, a higher level of granularity may be chosen. The goal is to achieve an actionable level where claims can be justified with assigned evidence.
 In the \textbf{Evidence block} of the CASCADE approach, security artifacts are provided to justify the arguments. Evidence can be provided at any stage of the argumentation process. For instance, in the black-hat block, if it can be proven that an asset is not susceptible to any threat scenario, evidence can be provided to justify the corresponding claims. The inability to assign evidence to claims indicates either that the argument has not reached an actionable point or that development changes are required to meet the claims. For example, if a claim lacks coverage in test reports, additional test cases may need to be created to address that claim.

 \textbf{Case Quality Assurance} in CASCADE incorporates quality assurance as per our design goal. It focuses on two key aspects.
 Firstly, \emph{completeness} ensures the coverage of claims in each argumentation level of the SAC. We introduce completeness claims for each strategy to refine and validate its coverage. These claims verify that the strategy includes all relevant claims on the argumentation level, considering the context provided.

 The second aspect, \emph{confidence}, measures the certainty of claim fulfillment based on the evidence provided. It is incorporated at each level of the security assurance case where at least one claim is justified by evidence. A confidence claim is formulated as follows: \emph{``The evidence provided for claim X achieves an acceptable level of confidence.''} The definition of an acceptable level of confidence is context-dependent and is established within the strategy. The confidence claim, like other claims, must be supported by evidence.

 To evaluate CASCADE, we collaborated with a cybersecurity expert from Volvo Trucks, a leading truck manufacturer in Sweden. Throughout the development process, we held multiple sessions with the expert to discuss the approach, its limitations, and potential improvements. In the next evaluation session, we presented CASCADE and a specific case related to headlamp items in accordance with ISO/SAE-21434. The expert assessed the approach by considering the overall structure of a SAC that would meet the security case requirements of ISO/SAE-21434 and aligned the elements of the example case with the company's internal practices. Valuable insights and suggestions for further enhancements were provided by the expert.
 As an additional validation step, we systematically created a mapping between the requirements and work products of ISO/SAE-21434 to elements of SAC and determined the corresponding blocks and levels of CASCADE. Requirements were mapped to elements such as assumptions, claims, case quality-claims, or case quality-evidence. Work products were classified as either context or evidence based on their scope and content.
 This helped us validate that SACs created with CASCADE have the capacity to include claims and evidence that cover all the requirements and work products of the standard.  
 


\begin{figure}
\begin{center}
  \includegraphics[width=1\linewidth]{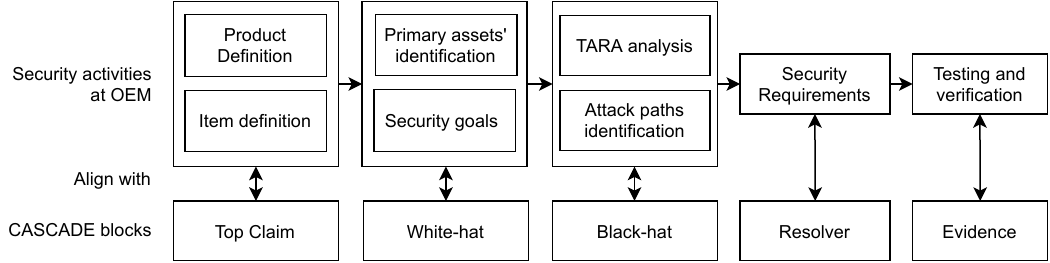}
\end{center}
        \caption{Mapping of the company's security activities to CASCADE blocks} 
        \label{fig:mappingKappa}
\end{figure}

 Figure \ref{fig:mappingKappa} depicts the mapping between various security activities at the company and the corresponding CASCADE blocks. The linkage signifies that the outputs of each activity are utilized to construct the SAC components in the related block. It can be observed from the figure that CASCADE is well-aligned with the company's operational approach.

\subsection{RQ4: To what extent can an approach for creating SAC in automotive be applied in other safety-critical domains?} 
 To answer this question, we conducted Paper D which is a case study at a large company in the medical domain.
 In the study, we identified several use cases for CASCADE in the medical domain. In terms of compliance, roles such as Device Regulator Lead, QA Owner, and Risk and Cybersecurity teams can use SAC created with CASCADE to demonstrate compliance with standards like FDA guidance and ISO 14971. Assessment-related use cases include determining the readiness for product release and assessing the quality of the product. The SACs were found to be useful for project planning and monitoring, including resource planning, identifying next steps, tracking the progress of suppliers, and ensuring full coverage of security requirements.

 The study also found significant overlaps between CASCADE and regulatory standards in the medical domain. The investigation covered various regulatory standards and guidance documents such as FDA, MDCG, EMA, ISO, and NIST. For example, ISO 14971, which focuses on risk management, aligns with CASCADE in terms of documentation, risk control, traceability, completeness of risk management, and personnel competence requirements. The study emphasized the importance of considering the relationship between safety and security, as security issues can have safety impacts.

 In summary, the study demonstrated the usefulness of CASCADE in the medical domain through identified use cases related to compliance, assessment, and project planning/monitoring. The study also highlighted the substantial overlap between CASCADE and regulatory standards, particularly in the context of risk management and safety considerations.

\subsection{RQ5: To what extent can the creation of security arguments be supported by a machine learning model for predicting security requirements?} 
Creating a SAC is not a trivial task and requires going through lots of documents, e.g., requirements specifications, to create the arguments. 
It is common that security-related requirements are not explicitly labeled in these documents. Hence, the creators of the cases would have to examine all requirements to determine the ones that need to be considered in the creation of the arguments. 
In Paper E, we created a machine-learning model to support the creation of security arguments by predicting if a requirement is security-related or not. We used data collected from 15 heterogeneous projects available online and tested it in a cross-project manner by training it on 14 documents and testing on the remaining one.

 The results of cross-project prediction indicate that training a classifier on other projects can be beneficial for the initial classification of a large set of requirements. 
 The classifier achieved an average precision above 84\% and an average recall of 77\%.
 The precision, recall, and f-measures achieved in this study are significantly better than those reported in previous research. To improve the classifier's performance, it is recommended to use training data that is heterogeneous in terms of phrasing, including specifications from different parts of the organization and written by various analysts and practitioners.

Furthermore, we tested the model on real-life regulatory documents taken from the automotive domain. We performed two sets of tests. The first was to predict if individual requirements are security related, while the second tests whether an entire section in the regulation includes at least one security-related requirement.

 The results for the performance of the classifier for individual security-related requirements in regulatory documents are shown in Table \ref{tab:results:regulations}. The results indicate that a classifier trained on different requirements documents can effectively predict security requirements. The aggregated section-based performance results demonstrate higher precision in identifying relevant sections in the regulations as shown in Table \ref{tab:results:sections}. However, there were instances of false positive predictions and missed security-related requirements, indicating areas for improvement in the classifier's performance. Overall, the classifier proves useful in predicting security requirements in regulatory documents, with the potential for further refinement.

  By analyzing the results we conclude that heterogeneity and the number of requirements in the training data were found to be significant factors for classification performance. High-quality labeling of training data and suitable training data with relevant security mechanisms and standards improved classification quality. The classifier can help prioritize documents and sections that require attention, aiding in the efficient processing of large amounts of information. However, misclassifications can occur due to specific terminology, the use of security standards without descriptions, and implicit references to security. False negatives were observed in certain instances, primarily due to differences in terminology and implicit references in the regulatory documents.

 To conclude, we believe that a machine-learning model can be beneficial in predicting security-related requirements and supports the creators of SAC in the creation of security arguments.


\begin{table}[tb]
\small
\centering
\caption{Classification of individual security-related \emph{requirements} in regulatory documents}
\label{tab:results:regulations}
\begin{tabular}{@{}p{1.4cm}p{1.2cm}p{1.2cm}p{0.8cm}p{0.5cm}p{0.6cm}p{0.6cm}@{}}
\toprule
\multicolumn{4}{l}{} & \multicolumn{3}{l}{\textbf{f-measures}} \\
\textbf{Regulation} & \textbf{Accuracy} & \textbf{Precision} & \textbf{Recall} & \textbf{f${}_1$} & \textbf{f${}_{1/2}$} & \textbf{f${}_2$}\\
\midrule
UN-R155 & 83.5 & 74.2 & 77 & 75.4 & 74,7 & 76,4 \\ 
UN-R156 & 99 & 100 & 98 & 94 & 99,6 & 98,4 \\ 
UN-R157  &  97.1  &  80  &  50  &  61.5  &  71,4  &  54 \\ 

\midrule
Avg.  & 93.2 & 84.7 & 72 & 77 & 76,6 & 76,3 \\ 

\bottomrule
\end{tabular}
\end{table}
\begin{table}[tb]
\small
\vspace{-2mm}
\centering
\caption{Classification of security-related \emph{sections} in standards UN-R155, UN-R156, and UN-R157.}
\label{tab:results:sections}
\begin{tabular}{@{}llllll@{}}
\toprule
\textbf{Standard} & \textbf{Precision} & \textbf{Recall} & \textbf{f${}_1$} & \textbf{f${}_{1/2}$} & \textbf{f${}_2$} \\
\midrule
UN-R155 & 100 & 71.4 & 83.3 &92.6&75.7\\
UN-R156 & 100 & 100 & 100 & 100 & 100\\
UN-R157 & 66.7 & 66.7 & 66.7 & 66.7 & 66.7\\

\bottomrule
\end{tabular}
\end{table}

\subsection{RQ6: How can security evidence be managed to support the creation of SACs?} 
To study the management of security evidence in safety-critical organizations, Paper F was conducted. In the study, six cases were included from three different domains. 
We found that organizations have immature practices in managing security evidence. The case companies are in the early stages of implementation and documentation of security evidence, with varying levels of maturity across different teams in their organizations. Drivers for working with security evidence include compliance, customer demands, market opportunities, and mitigating litigation risks. The study emphasizes the need for prioritizing security to protect organizations and customers, comply with regulations, and utilize evidence effectively.

Regarding how the management of security evidence is integrated into the development process of organizations, various activities throughout the development lifecycle generate different types of evidence. Examples of these activities are risk analysis, gap analysis, verification activities, security training, awareness programs, and incident logs. 

 Regarding the responsibility of evidence management, we identified five main activities and the roles responsible for carrying them out. These are shown in Table \ref{tbl:kappa:resp} 
 
 \begin{table*}
\caption{Roles and responsibilities of managing security evidence}
\label{tbl:kappa:resp}
\begin{center}
\resizebox{\textwidth}{!}{
\begin{tabular}{lccccc}
\toprule
             
            \textbf{Role} & \textbf{Creation}       & \textbf{Ownership}  &   \textbf{Collection}  &   \textbf{Maintenance}  & \textbf{Governance}  \\ 

\midrule
         Developer / DevOps   & \checkmark     &  &                &   \checkmark &        \\
         Product owner   &      &  \checkmark &    &   &    \checkmark     \\
         Risk owner   &     &   \checkmark &      & &         \\
         Auditor   &      &  &       &  & \checkmark          \\
         Management   &      &   &       & &  \checkmark       \\
         Security officer   & \checkmark   &   & \checkmark &   \checkmark   &\checkmark        \\
         Legal team & &   & \checkmark & & \\
             
\bottomrule
\end{tabular}}
\end{center}
\footnotesize
\end{table*}

 Storing the evidence can be done centrally with controlled access or in decentralized project/team repositories. It is structured based on features or architecture, and its quality is ensured through guidelines, audits, risk analysis, and metrics. 

 Managing security evidence presents significant challenges for organizations. These include establishing the scope amidst evolving threats, sharing sensitive data, addressing security throughout the product lifecycle, estimating cost and effort, overcoming competency gaps, and navigating vague guidelines in standards and regulations.

 Finally, the study discusses the significant benefits of automating tasks related to security evidence. Areas, where automation can be applied, include report generation, anomaly detection, testing and verification, and threat modeling. The industry's needs for automation involve security relevance prediction, test case generation and selection, traceability, change impact analysis, and threat modeling. Automation improves risk handling, evidence creation, and overall security practices, saving time and reducing errors.

 Based on the results of the study, we draw the following key insights:
 \begin{itemize}
     \item There is awareness of the importance of managing security evidence. However, there is a big gap between the current maturity levels and those needed to cope with the growing requirements.
     \item Many security artifacts are created throughout the development process. However, they are currently not considered evidence and are thus not covered by specific processes to manage them. Rather, they are managed as any other development artifact.
     \item Companies carry out many activities to manage development security evidence on a team level, but there is a lack of an organizational-level framework to manage evidence.
     \item The main challenges are organizational and related to structuring the work with security evidence and establishing the skill set rather than technical.
     \item There are interesting ideas for automation, but practitioners do not yet understand the capabilities of AI and how it can help.
     \item Effective management of security evidence requires considering the human aspect, including stakeholders’ needs and the human role in evidence quality assessment. Establishing a security culture is crucial, but there are challenges such as finding employees with security competencies and addressing the perception that work with evidence is overhead.
    \item The supplier-customer relationship in the context of evidence management is increasingly complex. A common  language for security assurance is key to coping with that, and automation can pave the way towards that
 \end{itemize}
\begin{figure}[!]
\begin{center}
  \includegraphics[width=\linewidth]{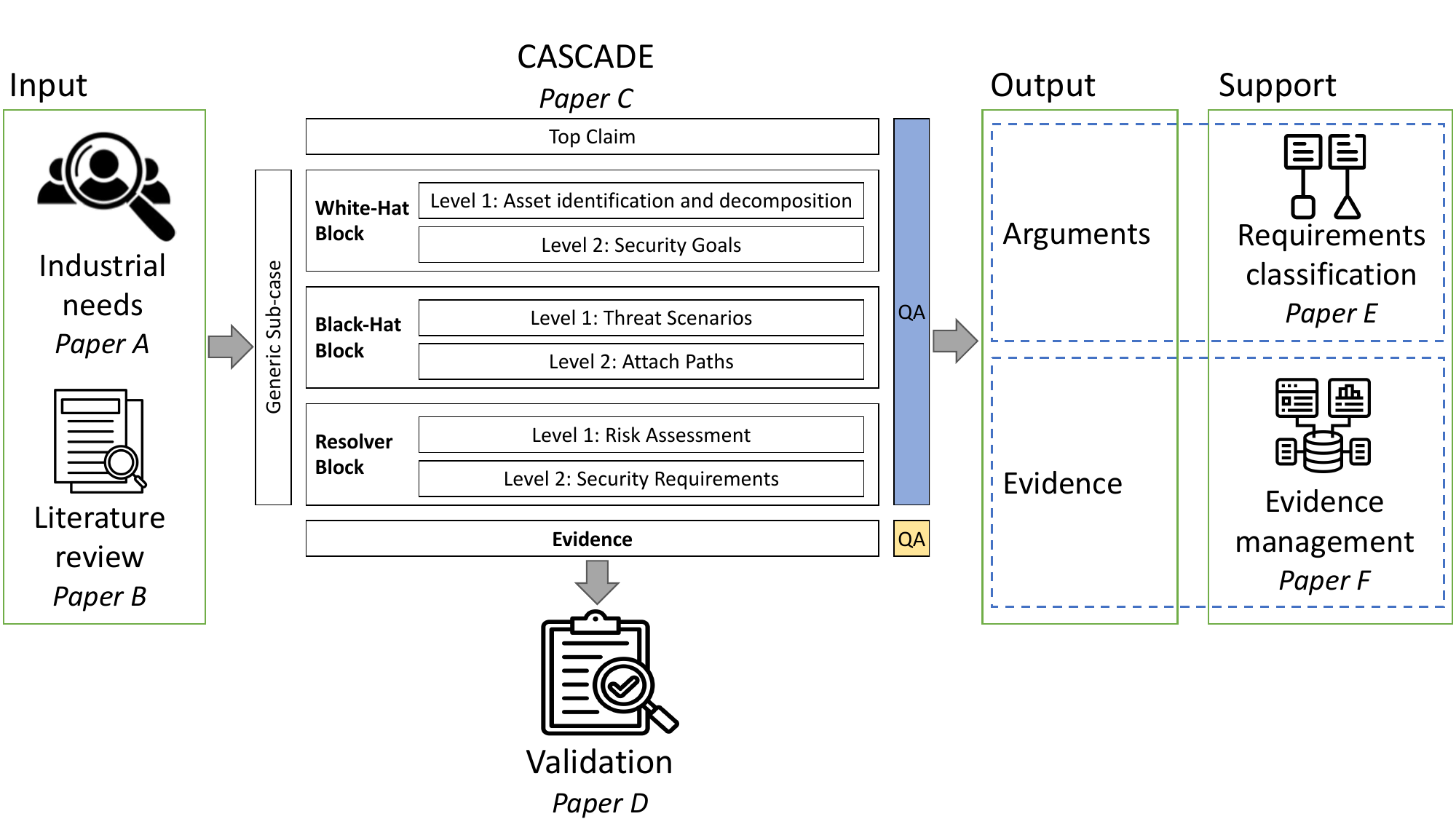}
  \caption{A summary of the contributions of this thesis} 
  \label{fig:kappa:contributions}
\end{center}
\end{figure}

To summarize, we illustrate the contributions of this thesis in Figure \ref{fig:kappa:contributions}.
As shown in the figure, we used a systematic literature review and a study of the industrial needs of SAC as inputs to create CASCADE, an approach for creating SAC. We validate the generalizability of CASCADE in another study and we support the output of CASCADE with a study for SAC argument creation and another for managing SAC evidence.





\newpage
\section{Threats to validity}

In this thesis, we consider the internal and external categories of validity threats as defined in \cite{ttv1}, and described in \cite{ttv3,kh2007guidelines}. 

In any case, the results presented in this paper are an important first important step towards a larger survey study involving more companies and professionals, internationally.
 With respect to \emph{external validity}, we acknowledge that the generalizability of our findings in Paper A and Paper C may be limited to the participating companies, which are from the same country. Therefore, our results may not be directly applicable to companies with different cultural backgrounds.
 Similarly, the findings in Paper F may not be applicable to all companies in safety-critical domains and markets due to variations in regulations, standards, and best practices across different domains. 
 However, the involved companies in the three studies are large and have high profiles. They are able to provide a quite broad perspective on the safety-critical domain as they also compete at the international level.
When it comes to the generalizability of our approach CASCADE, we dedicated a significant part of Paper D to study if it is applicable in the medical domain as it is for automotive.

 Regarding Paper E, we address external validity concerns related to machine learning models, i.e., overfitting and imbalanced data sets. We carefully split the data based on different projects from diverse sources to mitigate overfitting risks. Additionally, we used the synthetic minority oversampling technique to handle imbalanced data sets and ensure more reliable results beyond the specific data used in our study.


 Regarding \emph{internal validity}, we have identified several factors that require consideration. One potential limitation pertains to the prioritization of usage scenarios in Paper A, which may have been influenced by market pressures to comply with upcoming standards, potentially introducing bias in the selection of the top scenarios. Additionally, the selection of participants for the workshop and interviews in Paper A, as well as the evaluation in Paper C, was based on convenience sampling and expertise availability, which may also introduce limitations.
 However, we believe that the balance of participants in Paper A, representing diverse expertise in security, product development, business, and legal domains, as well as the experienced security expert involved in Paper C, provides us with confidence that the results are representative of the expectations and needs across the studied companies. Although the business knowledge around scenarios was prioritized, we acknowledge the potential limitations of these factors and aim to address them in future studies.



 The internal validity of the SLR conducted in Paper B could have been affected by subjectivity and the risk of missing relevant results since it was performed by only one researcher. To address this, a preliminary list of known good papers was manually created and other authors periodically performed quality checks. Additionally, publication bias could compromise the conclusion validity of the SLR, as studies with positive results are more likely to get published. 
To mitigate this, technical reports were also included in the review. The SLR is considered reliable, as any researcher with access to the used libraries can reproduce the study and obtain similar results. Furthermore, additional results may be obtained for studies published after the SLR's completion.

 Paper C utilized an example from ISO/SAE-21434 to demonstrate the use of CASCADE. Nevertheless, there is a potential that the example used may not accurately reflect real-world cases from the industry. Our assessment is that the structure of the example case is more significant for the evaluation than the specific content, which has been verified by the security expert who conducted the evaluation at the OEM.

In Paper E, the internal validity threats revolve around data labelling and algorithm selection. While the data was labelled by two individuals with relevant experience, there is a possibility of subjective judgement. To address this, a quality assurance step was performed. The selection of the Random Forest (RF) algorithm was based on a preliminary run, which may not necessarily be the optimal choice. Additionally, the study tested two feature extraction models, excluding recent innovations like BERT, with the goal of investigating cross-project prediction feasibility rather than finding the best approach.

In Paper F, the internal validity concerns are related to the thematic analysis process. To mitigate the risk of descriptive validity threat, the entire focus group sessions were recorded to ensure accurate documentation. Reactivity was minimized by promoting discussions among participants during the interviews. The transcribed sessions were sent back to participants for clarification, further reducing misunderstandings. For thematic coding, two researchers coded parts of the transcripts independently and discussed any discrepancies to establish a coding baseline. To analyze the codes and identify patterns, all four researchers collaborated in workshops, aiming to avoid subjectivity and researcher bias.

\section{Conclusion}

To conclude, the growing demand for connectivity in safety-critical domains has made security an increasingly important matter. As a result, regulatory and standardization bodies have mandated the use of Security Assurance Cases (SAC) to ensure the security of products and processes. In this context, CASCADE, an approach for creating SAC that aligns with ISO/SAE-21434 and incorporates quality assurance measures, has been developed. CASCADE addresses the gaps between the state-of-the-art in the literature and industrial needs for SAC of organizations in the safety-critical domain. We have demonstrated through an evaluation of CASCADE its suitability for integration into industrial product development processes both in the automotive and the medical domains. Furthermore, we have supported the creation and management of SACs by creating a machine-learning model to classify security-related requirements for use in the argumentation part of the SAC. 


\section{Future work}

SAC is a crucial area of research that has many interesting facets to explore. One of these areas involves examining how to maintain cases after their creation. For example, how to make them living documents that can be incorporated into a product's life-cycle. 

Another aspect involves studying how cases can change based on decisions made at run time. While cases are typically created during design time, many assumptions may not hold at run time. As a result, decisions made during that stage can impact the measures taken and potentially harm the security assurance case's integrity. One research topic that could be explored is incorporating the decision-making process into the arguments of an assurance case.

Currently, much of the work on SAC is driven by external factors, such as regulatory demands for proof of security. However, as the world becomes increasingly connected and the media frequently reports on various attacks, security will become a concern for everyone. Consequently, customers will likely start demanding proof of security for the products they purchase, such as various IoT devices. 
This creates a need for companies to find ways to provide security assurance to their customers. While SACs are simple structures, they can become very complicated when dealing with complex products. As a result, there is a possibility to use SAC toward end users, but also a need for research into how to abstract security arguments to a level that is easily understandable and communicable to customers, while also addressing the sensitivity of data.

\cleardoublepage 


\cleardoublepage\addcontentsline{toc}{chapter}{Bibliography}
\bibliographystyle{IEEEtran}

\bibliography{lit}

\end{document}